\documentclass[twocolumn,showpacs,preprintnumbers,amsmath,amssymb]{revtex4}
\usepackage[dvips]{graphicx} 

\newcommand{\cM}{{\mathcal M}}

\newcommand{\be}{\begin{equation}}
\newcommand{\ee}{\end{equation}}
\newcommand{\beq}{\begin{eqnarray}}
\newcommand{\eeq}{\end{eqnarray}}

\newcommand{\f}{\frac}
\newcommand{\vphi}{\varphi}

\begin{document}

\title{Entanglement in a multiverse with no common space-time}

\author{S. J. Robles-P\'{e}rez}

\address{F\'{\i}sica Te\'orica, Universidad del Pa\'{\i}s Vasco, Apartado 644, 48080, Bilbao (SPAIN).}

\begin{abstract}
Inter-universal entanglement may even exist in a multiverse in which there is no common space-time among the universes. In particular, the entanglement between the expanding and contracting branches of the universe might have observable consequences in the dynamical and thermodynamical properties of one single branch, making therefore testable the whole multiverse proposal, at least in principle.
\end{abstract}


\maketitle

\section{Introduction}

Entanglement is a quantum feature that provides us with a wide new variety of physical phenomena to be explored in cosmology too. Quantum entanglement among the modes of cosmic matter fields in a single-universe scenario shifts their ground state, having therefore distinguishable consequences in the energy levels \cite{PFGD1992a, Lee2007, Muller1995}, at least in principle. 

In some multiverse scenarios, particularly those in which the universes share a common space-time \cite{Mersini2008b}, the modes of the matter fields that belong to different universes could be entangled and it may thus induce observable imprints of the multiverse in a single universe \cite{Mersini2007, Mersini2008}. There is yet another multiverse scenario in which the universes share no common space-time \cite{Mersini2008b, Tegmark2003, RP2010}. In such multiverse the universes are completely disconnected from the point of view of the causal relations among the events that belong to different universes. Even though, quantum entanglement between the state of two or more universes might have observable consequences too \cite{RP2011}.

It is a challenging question how can entanglement be produced between two otherwise disconnected universes. In quantum mechanics, quantum entanglement between the state of two distant particles is present because at some previous time the two particles were interacting. More generally speaking, they formed part of a non-separable state that is preserved besides the later distance between the particles.

A multiverse scenario in which the different universes share no common space-time surely challenges our most fundamental notions in physics. In such a scenario, the universes may interact as the result of a residual interacting term from some dimensional reduction or compactification in a multi-dimensional theory. It can also be the result of a common origin in a classically forbidden region that may give rise to a pair of entangled universes \cite{PFGD1992a, RP2011}, or because a cosmic singularity splits the whole space-time manifold into two disconnected regions like it happens in a phantom dominated universe \cite{PFGD2007}. In any case, if the interaction among universes cannot be depicted in a common space-time, then, it becomes meaningless asking whether the universes interacted before or will interact after some given time, $t$.

For instance, it might well be that the universes could be created in entangled pairs from the Euclidean region of the space-time \cite{PFGD1992a, RP2011}. In that case, the quantum state of one single universe, i.e. the state that is obtained by tracing out the degrees of freedom of the partner universe, would depend on the properties of the other universe of the entangled pair because their properties are correlated from the very beginning. Thus, the entanglement between two or more universes provides us with a tool for testing the multiverse proposal, at least in principle.

In this letter, we present a developing description of the quantum entanglement among universes in the case of a multiverse in which there is no common space-time among the universes. In Sec. II, we address some general questions about the representation of universes in the third quantization formalism and its relation to the boundary conditions to be imposed on the state of the whole multiverse. In Sec. III, we analyze a possible scenario where inter-universal entanglement is contemplated and it is computed the thermodynamical magnitudes of entanglement. Finally, we draw some preliminary conclusions in Sec. IV.

\section{Quantum state of the multiverse in the third quantization formalism}

The basic idea of the third quantization formalism is to consider the Wheeler-deWitt equation as a Klein-Gordon equation \cite{McGuigan1988, Rubakov1988} and the wave function of the universe as the field to be quantized that propagates in the superspace. Thus, the state of the multiverse could be described within the general formalism of a quantum field theory of the superspace. Unfortunately, such an appealing approach can be properly defined only in minisuperspace models with a high degree of symmetry like the one that corresponds to a homogeneous and isotropic space-time. Nevertheless, a homogeneous and isotropic space-time is a well approximation for the description of the large parent universes that will populate our multiverse so we can certainly make use of the third quantization formalism.

The parallelism between the third quantization formalism and a quantum field theory has nevertheless obvious limits. For instance, the appropriate representation of particles in a quantum field theory is given, whenever it is possible, by the representation of particles that would be measured by a detector placed in an asymptotically flat region of the space-time, where particles can be well-defined.

In the multiverse, however, an observer can only perceive her own single universe which, on the other hand, may stay in an excited state rather than in the ground state \cite{Hartle1983}. Such an excited state would quantum mechanically be described by a quantum number, $| n \rangle$, that by no means can be identified with a 'number of universes' from the point of view of an internal observer. That concept would only become meaningful for an idealized 'super-observer', i.e. a detector that would live in the multiverse and that could thus measure different number of universes. 

In the case being considered here, such a hypothetical detector would not be defined in any space-time but in a more general abstract space where spatial and temporal relations are meaningless. In that case, the number of universes of the multiverse would presumably be a property that should not depend on the spatial or temporal properties of a particular single universe. It seems therefore appropriate to impose the boundary condition that the number of universes of the multiverse does not depend on the value of the scale factor of a particular single universe, at least from the point of view of the 'super-observer'. This boundary condition restricts the possible representations of universes in the multiverse to the set of invariant representations \cite{Lewis1969, Kim2001} under the change of a generic value of the scale factor, which plays the role of an intrinsic time variable in the multiverse. It has not to be confused with a time variable measurable by the clock of any observer who lives within a single universe. It is just the geometrical structure of the minisuperspace being considered that allows us to take the scale factor as the intrinsic time-like variable of the minisuperspace \cite{McGuigan1988, Rubakov1988}. The relationship between the multiverse and the arrow of time of a single universe, if any, should be analyzed \emph{a-posteriori}.

We arrive then at two significant representations: the 'invariant representation', which is the representation induced by a consistent boundary condition imposed on the state of the whole multiverse; and the 'asymptotic representation', which labels the excitation levels of a large parent universe from the point of view of an observer  inhabiting it.

\section{Thermodynamics of entanglement in the multiverse}

Specifically, let us consider a closed homogeneous and isotropic space-time endorsed with a cosmological constant, $\Lambda$, and a massless scalar field, $\varphi$, that would represent in a first approximation the matter content of the universe. In the third quantization formalism the wave function of the universe, $\phi\equiv\phi(a,\vphi)$, can be decomposed in normal modes, $\phi_k(a)$, with $\phi = \int dk \, e^{i k \varphi} \phi_k(a)$, that satisfy the Wheeler-deWitt equation. This can be written, in the model being considered, as \cite{RP2012b} 
\begin{equation}\label{WDW}
\ddot{\phi}_k + \f{\dot{\cM}(a)}{\cM(a)} \dot{\phi}_k + \omega^2_k(a) \phi_k = 0 ,
\end{equation}
where, $\mathcal{M}(a) \equiv a^p$ depends on the choice of factor ordering (the customary choice \cite{Kiefer2007} corresponds to the value $p=1$), the dot means the derivative with respect to the scale factor, $a$, and
\be\label{frequency01}
\omega_k^2(a) = \Lambda a^4 - a^2 + \frac{k^2}{a^2}.
\ee
The invariant representations of the harmonic oscillator are well-known \cite{Lewis1969, Kim2001}. They are described in terms of creation and annihilation operators, $\hat{b}_k^\dag\equiv\hat{b}_k^\dag(a) $ and $\hat{b}_k\equiv\hat{b}_k(a)$, which  are invariant under the action of the Hamiltonian that gives rise to Eq. (\ref{WDW}) through an appropriate generalization of the Heisenberg equations. Here, it corresponds to the Hamiltonian of a harmonic oscillator with mass $\mathcal{M}(a)$ and frequency $\omega_k(a)$. As it is also well-known, the different representations of the harmonic oscillator are related to each other by a squeezed transformation \cite{Vedral2006}. The invariant operators, $\hat{b}_k^\dag$ and $\hat{b}_k$, are thus related to the creation and annihilation operators of the asymptotic representation, $\hat{c}_k^\dag$ and $\hat{c}_k$, that correspond to the ladder operators of the excited levels of the universe as being perceived by  an internal observer, i.e. $\hat{b}_k = \alpha_k \, \hat{c}_k + \beta_k \, \hat{c}_k^\dag$, with $|\alpha_k|^2-|\beta_k|^2 = 1$.

Let us assume that the multiverse stays in the ground state of the 'super-observer' representation, $|_{(b)} 0_{k,-k}\rangle$, of a particular single mode $k$. It can be written in the 'observer' representation, $|_{(c)} n_{k}, n_{-k}\rangle$, as \cite{Mukhanov2007}
\begin{equation}
|_{(b)} 0_{k,-k}\rangle = \frac{1}{|\alpha_k|} \sum_{n=0}^\infty \left( \frac{\beta_k}{\alpha_k} \right)^n |_{(c)} n_{k}, n_{-k}\rangle ,
\end{equation}
where the $-$ and $+$ signs of the $k$-modes correspond here to the contracting and expanding branches of the universe \cite{RP2011}, respectively. The density matrix that represents the composite quantum state is then given by $\rho_k \equiv |_{(b)} 0_{k,-k}\rangle \langle _{(b)}0_{k,-k}|$, and the reduced density matrix that represents the quantum state of one single branch of the universe by \cite{RP2011, RP2012b}
\begin{equation}\label{ts}
\rho_k \equiv {\rm Tr}_{-k} \rho = Z^{-1}  \sum_{n=0}^\infty e^{-\frac{\omega}{T} (n+\frac{1}{2})}  |_{(c)}n_k\rangle\langle _{(c)}n_k| ,
\end{equation}
with, $Z = |\alpha_k| |\beta_k|$. Eq. (\ref{ts}) describes a thermal state with a generalized temperature, $T \equiv T(a) \equiv \frac{\omega_k(a)}{2 ( \ln |\alpha_k| - \ln |\beta_k| )}>0$, that depends on the value of the scale factor. It is now straightforward to compute the thermodynamical magnitudes \cite{Alicki2004} associated to the thermal state (\ref{ts}). However, we should notice that these are thermodynamical magnitudes of entanglement in a multiverse scenario that parallels but generalizes the space-time scenario in which the customary magnitudes of classical thermodynamics are formulated. It may be therefore not surprising that the mean value of the Hamiltonian $H_k \equiv \omega_k (\hat{c}_k^\dag \hat{c}_k + \frac{1}{2})$ does not correspond to the value of the energy density of one single universe, $\varepsilon(a)$, which is effectively given by Eq. (\ref{frequency01}), with $a^4 \varepsilon(a) = \omega_k^2(a)$ [let us recall that classically, $\omega_k(a) \equiv a \partial_ta$, where $t$ is the cosmic time]. The mean value, $\langle H_k \rangle \equiv {\rm Tr}_k(\rho_k H_k)$, is given by 
\begin{equation}
\label{energy}
E = \omega_k (\langle n_k \rangle +\frac{1}{2}) ,
\end{equation}
with, $\langle n_k \rangle \equiv |\beta_k|^2$. Then, the entanglement between two universes would presumably contribute to the gravitational energy density of a single universe with a new term in Eq. (\ref{frequency01}) that would account for the relation (\ref{energy}) and, in principle, it ought to have observable consequences on the dynamical and thermodynamical properties of that universe. It is expected, however, that the effect of inter-universal entanglement is too small except may be for the very early stage of the universe, where the effect is stronger \cite{RP2010, RP2012}.

The entropy of entanglement, $S_{ent}$, given by the von-Neumann formula applied to the density matrix (\ref{ts}), turns out to be a decreasing function with respect to an increasing value of the scale factor \cite{RP2011, RP2012}. However, the second principle of thermodynamics is still satisfied because the evolution of each single universe is not adiabatic, in the quantum informational sense, and the heat decreases as the universe expands. More exactly, the production of entropy $\sigma$, defined as \cite{Alicki2004} $\sigma \equiv dS_{ent} - \frac{\delta Q}{T}$, is zero and
\begin{equation}
dS_{ent} = \frac{\delta Q}{T} < 0 .
\end{equation}
It is worth noticing that, $dS_{ent} = \frac{\delta Q}{T} > 0$, for the contracting branch of the universe. These results do not contradict those previously obtained in Refs. \cite{Hawking1985, Hawking1992, Page1985, Kiefer1995} because we are dealing here with the entropy of entanglement that, in principle, it is completely different from the entropy that corresponds to the inhomogeneous degrees of freedom of the matter fields of the universe. In the context of the multiverse the universe needs not to be a closed system and, indeed, the expansion or the contraction of the universe is not an adiabatic processes in the model being considered here (in the quantum informational sense) because there is a change of the rate of entanglement between the two universes of the entangled pair. That is to say, the second principle of thermodynamics, $\sigma \geq 0$, is here satisfied for both the expanding and the contracting branches of the universe because \cite{RP2011b, RP2012, RP2012b}, $\sigma = 0$. The relation that may exist, if any, between the entropy of entanglement among universal states and the entropy that corresponds to the inhomogeneous degrees of freedom of the matter fields within a single universe scenario is a subject that deserves a deeper understanding of both the inter-universal entanglement and the relation between the classical formulation of thermodynamics and the thermodynamics of entanglement  \cite{Vedral2002, Brandao2008}.

\section{Conclusions}

Entanglement in a multiverse scenario is a novel feature that provides us with new cosmic phenomena to be explore. Furthermore, it might have observable and distinguishable consequences in the properties of our single universe and, thus, it might supply us with a way to test the whole multiverse proposal, at least in principle.

In this communication, we have presented a developing description of quantum entanglement in the context of a multiverse in which there is no common space-time among the universes of the multiverse. They are therefore causally disconnected from a classical point of view although their composite state may still show quantum correlations.

Two significant representations have been considered: the 'invariant representation', which is derived from the boundary condition imposed on the state of the whole multiverse, and the 'asymptotic representation' that describes the universe as it would be seen by an internal observer. They are both related by a Bogolyubov transformation and the quantum state of one single branch of an entangled pair of branches of the universe, one expanding branch and one contracting branch, turns out to be described by a thermal state with a value of the generalized temperature that depends on the scale factor.

The thermodynamical properties of entanglement can then be computed. The energy of entanglement should contribute to the energy content of each single universe, making therefore testable the multiverse proposal. However, it is still left the development of a well-defined relationship between the energy of inter-universal entanglement and the energy density of each single universe. That would allow us to analyze the observable consequences and to pose measurable tests.

The entropy of entanglement turns out to be a decreasing function with respect to an increasing value of the scale factor. However, the second principle of thermodynamics is still satisfied because the expansion or contraction of the universe is not an adiabatic process, in the quantum informational sense, and there is a change in the rate of entanglement as the universe expands or contracts. In the contracting branch the entropy of entanglement increases as the universe collapses. It would seem to be a result that supports other previous results \cite{Hawking1985, Hawking1992, Page1985, Kiefer1995}. However, it is not clear at all the relation that may exist, if any, between the the entropy of inter-universal entanglement and the entropy that corresponds to the inhomogeneous modes of the matter fields of a single universe. This is a subject that deserves further attention.

\section*{Acknowledgments}
This work was supported by the Basque Government project IT-221-07.

\bibliographystyle{apsrev}
\bibliography{bibliography}

\begin{thebibliography}{28}
\expandafter\ifx\csname natexlab\endcsname\relax\def\natexlab#1{#1}\fi
\expandafter\ifx\csname bibnamefont\endcsname\relax
  \def\bibnamefont#1{#1}\fi
\expandafter\ifx\csname bibfnamefont\endcsname\relax
  \def\bibfnamefont#1{#1}\fi
\expandafter\ifx\csname citenamefont\endcsname\relax
  \def\citenamefont#1{#1}\fi
\expandafter\ifx\csname url\endcsname\relax
  \def\url#1{\texttt{#1}}\fi
\expandafter\ifx\csname urlprefix\endcsname\relax\def\urlprefix{URL }\fi
\providecommand{\bibinfo}[2]{#2}
\providecommand{\eprint}[2][]{\url{#2}}

\bibitem[{\citenamefont{Gonz{\'a}lez-D{\'\i}az}(1992)}]{PFGD1992a}
\bibinfo{author}{\bibfnamefont{P.~F.} \bibnamefont{Gonz{\'a}lez-D{\'\i}az}},
  \bibinfo{journal}{Phys. Lett. B} \textbf{\bibinfo{volume}{293}},
  \bibinfo{pages}{294} (\bibinfo{year}{1992}).

\bibitem[{\citenamefont{Lee et~al.}(2007)}]{Lee2007}
\bibinfo{author}{\bibfnamefont{J.-W.} \bibnamefont{Lee}} \bibnamefont{et~al.},
  \bibinfo{journal}{JCAP} \textbf{\bibinfo{volume}{0708}}, \bibinfo{pages}{005}
  (\bibinfo{year}{2007}), \eprint{hep-th/0701199}.

\bibitem[{\citenamefont{M{\"u}ller and Lousto}(1995)}]{Muller1995}
\bibinfo{author}{\bibfnamefont{R.}~\bibnamefont{M{\"u}ller}} \bibnamefont{and}
  \bibinfo{author}{\bibfnamefont{C.~O.} \bibnamefont{Lousto}},
  \bibinfo{journal}{Phys. Rev. D} \textbf{\bibinfo{volume}{52}},
  \bibinfo{pages}{4512} (\bibinfo{year}{1995}).

\bibitem[{\citenamefont{Mersini-Houghton}(2008)}]{Mersini2008b}
\bibinfo{author}{\bibfnamefont{L.}~\bibnamefont{Mersini-Houghton}}
  (\bibinfo{year}{2008}), \eprint{arXiv:0804.4280v1}.

\bibitem[{\citenamefont{Mersini-Houghton}(2007)}]{Mersini2007}
\bibinfo{author}{\bibfnamefont{L.}~\bibnamefont{Mersini-Houghton}},
  \bibinfo{journal}{New Scientist} pp. \bibinfo{pages}{11--24}
  (\bibinfo{year}{2007}).

\bibitem[{\citenamefont{Holman et~al.}(2008)\citenamefont{Holman,
  Mersini-Houghton, and Takahashi}}]{Mersini2008}
\bibinfo{author}{\bibfnamefont{R.}~\bibnamefont{Holman}},
  \bibinfo{author}{\bibfnamefont{L.}~\bibnamefont{Mersini-Houghton}},
  \bibnamefont{and}
  \bibinfo{author}{\bibfnamefont{T.}~\bibnamefont{Takahashi}},
  \bibinfo{journal}{Phys. Rev. D} \textbf{\bibinfo{volume}{77}},
  \bibinfo{pages}{063510,063511} (\bibinfo{year}{2008}),
  \eprint{[arXiv:hep-th/0611223v1], [arXiv:hep-th/0612142v1]}.

\bibitem[{\citenamefont{Tegmark}(2003)}]{Tegmark2003}
\bibinfo{author}{\bibfnamefont{M.}~\bibnamefont{Tegmark}},
  \bibinfo{journal}{Scientific American} \textbf{\bibinfo{volume}{288}}
  (\bibinfo{year}{2003}).

\bibitem[{\citenamefont{Robles-P{\'e}rez and
  Gonz{\'a}lez-D{\'\i}az}(2010)}]{RP2010}
\bibinfo{author}{\bibfnamefont{S.}~\bibnamefont{Robles-P{\'e}rez}}
  \bibnamefont{and} \bibinfo{author}{\bibfnamefont{P.~F.}
  \bibnamefont{Gonz{\'a}lez-D{\'\i}az}}, \bibinfo{journal}{Phys. Rev. D}
  \textbf{\bibinfo{volume}{81}}, \bibinfo{pages}{083529}
  (\bibinfo{year}{2010}), \eprint{arXiv:1005.2147v1}.

\bibitem[{\citenamefont{Robles-P{\'e}rez
  et~al.}(2011)\citenamefont{Robles-P{\'e}rez, Alonso-Serrano, and
  Gonz{\'a}lez-D{\'\i}az}}]{RP2011}
\bibinfo{author}{\bibfnamefont{S.}~\bibnamefont{Robles-P{\'e}rez}},
  \bibinfo{author}{\bibfnamefont{A.}~\bibnamefont{Alonso-Serrano}},
  \bibnamefont{and} \bibinfo{author}{\bibfnamefont{P.~F.}
  \bibnamefont{Gonz{\'a}lez-D{\'\i}az}}, \bibinfo{journal}{Phys. Rev. D}
  \textbf{\bibinfo{volume}{85}}, \bibinfo{pages}{063611}
  (\bibinfo{year}{2011}), \eprint{arXiv:1111.3178}.

\bibitem[{\citenamefont{Gonz{\'a}lez-D{\'\i}az and
  Robles-P{\'e}rez}(2008)}]{PFGD2007}
\bibinfo{author}{\bibfnamefont{P.~F.} \bibnamefont{Gonz{\'a}lez-D{\'\i}az}}
  \bibnamefont{and}
  \bibinfo{author}{\bibfnamefont{S.}~\bibnamefont{Robles-P{\'e}rez}},
  \bibinfo{journal}{Int. J. Mod. Phys. D} \textbf{\bibinfo{volume}{17}},
  \bibinfo{pages}{1213} (\bibinfo{year}{2008}), \eprint{0709.4038}.

\bibitem[{\citenamefont{McGuigan}(1988)}]{McGuigan1988}
\bibinfo{author}{\bibfnamefont{M.}~\bibnamefont{McGuigan}},
  \bibinfo{journal}{Phys. Rev. D} \textbf{\bibinfo{volume}{38}},
  \bibinfo{pages}{3031} (\bibinfo{year}{1988}).

\bibitem[{\citenamefont{Rubakov}(1988)}]{Rubakov1988}
\bibinfo{author}{\bibfnamefont{V.~A.} \bibnamefont{Rubakov}},
  \bibinfo{journal}{Phys. Lett. B} \textbf{\bibinfo{volume}{214}},
  \bibinfo{pages}{503} (\bibinfo{year}{1988}).

\bibitem[{\citenamefont{Hartle and Hawking}(1983)}]{Hartle1983}
\bibinfo{author}{\bibfnamefont{J.~B.} \bibnamefont{Hartle}} \bibnamefont{and}
  \bibinfo{author}{\bibfnamefont{S.~W.} \bibnamefont{Hawking}},
  \bibinfo{journal}{Phys. Rev. D} \textbf{\bibinfo{volume}{28}},
  \bibinfo{pages}{2960} (\bibinfo{year}{1983}).

\bibitem[{\citenamefont{Lewis and Riesenfeld}(1969)}]{Lewis1969}
\bibinfo{author}{\bibfnamefont{H.~R.} \bibnamefont{Lewis}} \bibnamefont{and}
  \bibinfo{author}{\bibfnamefont{W.~B.} \bibnamefont{Riesenfeld}},
  \bibinfo{journal}{J. Math. Phys.} \textbf{\bibinfo{volume}{10}},
  \bibinfo{pages}{1458} (\bibinfo{year}{1969}).

\bibitem[{\citenamefont{Kim and Page}(2001)}]{Kim2001}
\bibinfo{author}{\bibfnamefont{S.~P.} \bibnamefont{Kim}} \bibnamefont{and}
  \bibinfo{author}{\bibfnamefont{D.~N.} \bibnamefont{Page}},
  \bibinfo{journal}{Phys. Rev. A} \textbf{\bibinfo{volume}{64}},
  \bibinfo{pages}{012104} (\bibinfo{year}{2001}).

\bibitem[{\citenamefont{Robles-P{\'e}rez}(2012{\natexlab{a}})}]{RP2012b}
\bibinfo{author}{\bibfnamefont{S.~J.} \bibnamefont{Robles-P{\'e}rez}},
  \emph{\bibinfo{title}{Open questions in cosmology}}
  (\bibinfo{publisher}{InTech}, \bibinfo{year}{2012}{\natexlab{a}}), chap.
  \bibinfo{chapter}{Inter-universal entanglement}.

\bibitem[{\citenamefont{Kiefer}(2007)}]{Kiefer2007}
\bibinfo{author}{\bibfnamefont{C.}~\bibnamefont{Kiefer}},
  \emph{\bibinfo{title}{Quantum gravity}} (\bibinfo{publisher}{Oxford
  University Press, Oxford, UK}, \bibinfo{year}{2007}).

\bibitem[{\citenamefont{Vedral}(2006)}]{Vedral2006}
\bibinfo{author}{\bibfnamefont{V.}~\bibnamefont{Vedral}},
  \emph{\bibinfo{title}{Introduction to quantum information science}}
  (\bibinfo{publisher}{Oxford University Press, Oxford, UK},
  \bibinfo{year}{2006}).

\bibitem[{\citenamefont{Mukhanov and Winitzki}(2007)}]{Mukhanov2007}
\bibinfo{author}{\bibfnamefont{V.~F.} \bibnamefont{Mukhanov}} \bibnamefont{and}
  \bibinfo{author}{\bibfnamefont{S.}~\bibnamefont{Winitzki}},
  \emph{\bibinfo{title}{Quantum Effects in Gravity}}
  (\bibinfo{publisher}{Cambridge University Press, Cambridge, UK},
  \bibinfo{year}{2007}).

\bibitem[{\citenamefont{Alicki et~al.}(2004)}]{Alicki2004}
\bibinfo{author}{\bibfnamefont{R.}~\bibnamefont{Alicki}} \bibnamefont{et~al.},
  \bibinfo{journal}{Open Syst. Inf. Dyn.} \textbf{\bibinfo{volume}{11}},
  \bibinfo{pages}{205} (\bibinfo{year}{2004}),
  \eprint{arXiv:quant-ph/0402012v2}.

\bibitem[{\citenamefont{Robles-P{\'e}rez}(2012{\natexlab{b}})}]{RP2012}
\bibinfo{author}{\bibfnamefont{S.}~\bibnamefont{Robles-P{\'e}rez}},
  \bibinfo{journal}{(submitted)}  (\bibinfo{year}{2012}{\natexlab{b}}),
  \eprint{arXiv:1203.5774}.

\bibitem[{\citenamefont{Hawking}(1985)}]{Hawking1985}
\bibinfo{author}{\bibfnamefont{S.~W.} \bibnamefont{Hawking}},
  \bibinfo{journal}{Phys. Rev. D} \textbf{\bibinfo{volume}{32}},
  \bibinfo{pages}{2489} (\bibinfo{year}{1985}).

\bibitem[{\citenamefont{Hawking et~al.}(1992)\citenamefont{Hawking, Laflamme,
  and Lyons}}]{Hawking1992}
\bibinfo{author}{\bibfnamefont{S.~W.} \bibnamefont{Hawking}},
  \bibinfo{author}{\bibfnamefont{R.}~\bibnamefont{Laflamme}}, \bibnamefont{and}
  \bibinfo{author}{\bibfnamefont{G.~W.} \bibnamefont{Lyons}},
  \bibinfo{journal}{Phys. Rev. D} pp. \bibinfo{pages}{1546--1550}
  (\bibinfo{year}{1992}).

\bibitem[{\citenamefont{Page}(1985)}]{Page1985}
\bibinfo{author}{\bibfnamefont{D.~N.} \bibnamefont{Page}},
  \bibinfo{journal}{Phys. Rev. D} \textbf{\bibinfo{volume}{32}},
  \bibinfo{pages}{2496} (\bibinfo{year}{1985}).

\bibitem[{\citenamefont{Kiefer and Zeh}(1995)}]{Kiefer1995}
\bibinfo{author}{\bibfnamefont{C.}~\bibnamefont{Kiefer}} \bibnamefont{and}
  \bibinfo{author}{\bibfnamefont{H.~D.} \bibnamefont{Zeh}},
  \bibinfo{journal}{Phys. Rev. D} \textbf{\bibinfo{volume}{51}},
  \bibinfo{pages}{4145} (\bibinfo{year}{1995}).

\bibitem[{\citenamefont{Robles-P{\'e}rez and
  Gonz{\'a}lez-D{\'\i}az}(2011)}]{RP2011b}
\bibinfo{author}{\bibfnamefont{S.}~\bibnamefont{Robles-P{\'e}rez}}
  \bibnamefont{and} \bibinfo{author}{\bibfnamefont{P.~F.}
  \bibnamefont{Gonz{\'a}lez-D{\'\i}az}} (\bibinfo{year}{2011}),
  \eprint{arXiv:1111.4128}.

\bibitem[{\citenamefont{Vedral and Kashefi}(2002)}]{Vedral2002}
\bibinfo{author}{\bibfnamefont{V.}~\bibnamefont{Vedral}} \bibnamefont{and}
  \bibinfo{author}{\bibfnamefont{E.}~\bibnamefont{Kashefi}},
  \bibinfo{journal}{Phys. Rev. Lett.} \textbf{\bibinfo{volume}{89}},
  \bibinfo{pages}{037903} (\bibinfo{year}{2002}).

\bibitem[{\citenamefont{Brandao and Plenio}(2008)}]{Brandao2008}
\bibinfo{author}{\bibfnamefont{F.~G. S.~L.} \bibnamefont{Brandao}}
  \bibnamefont{and} \bibinfo{author}{\bibfnamefont{M.~B.}
  \bibnamefont{Plenio}}, \bibinfo{journal}{Nature Physics}
  \textbf{\bibinfo{volume}{4}}, \bibinfo{pages}{873} (\bibinfo{year}{2008}).

\end{thebibliography}

\end{document}